\documentclass[aps,pre,floatfix,superscriptaddress,showpacs,nofootinbib,amsmath,amssymb,reprint]{revtex4-2}
\usepackage[utf8]{inputenc}
\usepackage{float}
\usepackage[caption = false]{subfig}
\usepackage{amsmath}
\usepackage{amssymb}
\usepackage{graphicx}
\usepackage{verbatim}
\usepackage{xcolor}
\usepackage{epstopdf}
\usepackage{bbm}
\usepackage{url}
\usepackage{tikz}
\usepackage{comment}
\usepackage[pdftex]{hyperref}
\usepackage[normalem]{ulem}

\begin{document}
\title{Comparison of wave-mixing processes in rarefied gas and QED vacuum using numerical simulations}

\author{Marianna Lytova}
\email[]{marianna.lytova@inrs.ca}
\affiliation{Centre \'Energie Mat\'eriaux T\'el\'ecommunications, Institut national de la recherche scientifique (INRS), 1650 blvd. Lionel-Boulet, Varennes, J3X 1P7, QC, Canada}

\author{Fran\c{c}ois Fillion-Gourdeau}
\affiliation{Centre \'Energie Mat\'eriaux T\'el\'ecommunications, Institut national de la recherche scientifique (INRS), 1650 blvd. Lionel-Boulet, Varennes, J3X 1P7, QC, Canada}
\affiliation{Infinite Potential Laboratories, LP, 485 Wes Graham Way, Waterloo, N2L 6R2, ON, Canada}

\author{Simon Valli\`eres}
\affiliation{Centre \'Energie Mat\'eriaux T\'el\'ecommunications, Institut national de la recherche scientifique (INRS), 1650 blvd. Lionel-Boulet, Varennes, J3X 1P7, QC, Canada}

\author{Sylvain Fourmaux}
\affiliation{Centre \'Energie Mat\'eriaux T\'el\'ecommunications, Institut national de la recherche scientifique (INRS), 1650 blvd. Lionel-Boulet, Varennes, J3X 1P7, QC, Canada}

\author{Fran\c{c}ois L\'egar\'e}
\affiliation{Centre \'Energie Mat\'eriaux T\'el\'ecommunications, Institut national de la recherche scientifique (INRS), 1650 blvd. Lionel-Boulet, Varennes, J3X 1P7, QC, Canada}

\author{Steve MacLean}
\affiliation{Centre \'Energie Mat\'eriaux T\'el\'ecommunications, Institut national de la recherche scientifique (INRS), 1650 blvd. Lionel-Boulet, Varennes, J3X 1P7, QC, Canada}
\affiliation{Infinite Potential Laboratories, LP, 485 Wes Graham Way, Waterloo, N2L 6R2, ON, Canada}

\date{\today}

\begin{abstract}
We study the conditions required to distinguish laser-induced nonlinear quantum electrodynamics (QED) effects in vacuum from competing signals due to interactions of laser pulses with ionized residual gas. The latter is inevitably present in vacuum chambers where experiments are performed because the vacuum is never perfect and there is always some residual pressure. The rarefied gas contribution is modeled statistically using the 1D–1V Vlasov–Maxwell system, while vacuum nonlinearities are described within the weak-field expansion of the Heisenberg–Euler effective action. In both cases, photon spectra from wave-mixing processes are evaluated by solving numerically the resulting partial differential equations using a semi-Lagrangian scheme.  We consider short pulses in co- and counter-propagating configurations, allowing us to identify the laser intensities and vacuum pressures for which the vacuum signal dominates. These results provide quantitative guidance for future all-optical experiments aiming to detect light-by-light scattering in vacuum.

\end{abstract}

\maketitle


\section{Introduction}
Wave-mixing processes (WMP) are well-known phenomena in nonlinear optics \cite{Boyd}: the interaction of high power laser pulses can lead to the generation of new fields with frequencies that, under phase-matching conditions, may differ from the frequencies of the incident pulses. The physical nature of the nonlinearity underlying WMP depends on the properties of the medium considered. For example, in non-ionized gases, the nonlinear response is due to the polarization of bound charges \cite{Boyd81, Boyd}, while in plasmas it is caused by currents of free charged particles \cite{Steel:79, Ma}.

At ultra-high intensity, quantum electrodynamics (QED) predicts that the vacuum can be polarized via the creation of virtual electron-positron pairs. These pairs induce an effective dipole moment which in turns modifies the propagation of electromagnetic waves \cite{Euler1935,Heisenberg,Schwinger,Sauter,Rozanov}. For laser field strengths below the critical Schwinger value $E_{S}\approx1.32\cdot10^{18}$~V/m, this leads to effects akin to nonlinear phenomena in matter \cite{0034-4885-76-1-016401}, which include WMP and birefringence as specific examples. Although the latter phenomena are pure QED effects, their theoretical description in the perturbative regime has much in common mathematically with higher-order processes in nonlinear optics. In particular, it is reasonable to associate the four-wave mixing (FWM) process with an effective third-order susceptibility \cite{Moulin99, Steel:79, king_heinzl_2016}. Thus, just like classical nonlinear processes, it is possible that FWM radiation from vacuum has different spectrum, polarization and propagation direction than incoming laser fields, allowing us to distinguish them. For this reason, FWM has been proposed theoretically as an all-optical probe for QED effects and light-by-light scattering in vacuum \cite{particles3010005,Bohl,Shukla,1367-2630-14-10-103002,king_heinzl_2016,PhysRevA.74.043821,PhysRevD.103.076009,PhysRevD.97.076002,PhysRevD.93.125032}.  

Detecting QED nonlinear effects due to light-to-light scattering is a very challenging task because the interaction strength is weak, of $O(\alpha^{2})$ (to lowest order), where $\alpha \approx 1/137$ is the fine structure constant. Therefore, even with very high field strengths, the number of generated photons remains limited and hard to detect. Current advanced laser technologies are able to reach intensities up to $10^{22} - 10^{23}$~W/cm$^2$ \cite{Pirozhkov:17, Yoon:21}, which makes it possible, in principle, to study QED effects in vacuum in the perturbative and even in the non-perturbative regime (see e.g. Refs. \cite{Baumann, Shukla, Doyle_2022, Roso_2022} and review \cite{Fedotov}). However, no experiment has successfully detected light-by-light scattering using direct laser fields so far (in the optical or near-infrared regimes), although there are some theoretical suggestions on how this effect could be observed \cite{king_heinzl_2016,particles3010005,Shukla,PhysRevD.109.096009,Ahmadiniaz_2025} (see further Section~\ref{num_exp}) and there has been some experimental attempts \cite{moulin1996photon,moulin2000,app7070671,YAMAJI2016454,WATT2025139247,smid2025darkfield}. Conversely, the analogous phenomenon of light-by-light scattering from initial virtual high-energy photons has been observed experimentally in heavy ion collisions, providing stringent constraints on the cross-section at high frequencies \cite{atlas2017evidence,PhysRevLett.123.052001}. In all cases, the observable signal for QED effects is weak and could be easily mixed up with competing processes from the residual gas in the vacuum chamber where the experiment is performed. This background signal has been assessed for a specific experimental configuration in Ref. \cite{Doyle_2022}.

Although vacuum technology has made great strides in recent years, so that the pressure can be reduced down to $10^{-12}$ Torr with standard techniques \cite{weston2013ultrahigh}, the quality of ultra-high vacuum is never perfect because there is always some residual gas particles. Even with the degree of rarefaction that can be achieved and the actual intensity of laser pulses, it is possible that different types of processes causing WMP can be excited simultaneously. To study only QED effects, we need to determine under what conditions one can expect the predominance of WMP in vacuum over WMP from gas particles. 
Here, we employ numerical simulations to answer this question.

This article is organized as follows. Section~\ref{num_exp} provides general information on our numerical experiments. Detailed information on computational models of the WMP process in vacuum and from residual particles is presented in Sections~\ref{plasma} and \ref{vacuum}, respectively. The corresponding numerical algorithms are described in Appendices~\ref{plasma_alg} and \ref{vacuum_alg}. Section~\ref{res} presents and compares the results obtained for gas and vacuum. Section~\ref{cncls} concludes the paper. 

\section{Numerical experiment setup}\label{num_exp}
We start with a discussion of the interaction region geometry and its reasonable simplification for numerical modeling. Many possible experimental scenarios have been proposed for studying light-to-light scattering with laser pulses. Most of them use either the tight focusing of a single beam  \cite{Fillion,PhysRevLett.107.073602}, or the crossing of two \cite{1367-2630-14-10-103002,PhysRevA.82.032114,PhysRevA.82.011803,PhysRevA.86.033810,1367-2630-17-4-043060,Shukla,Roso_2022} and three \cite{Rozanov,Moulin99,moulin2000,Lund,PhysRevA.74.043821,PhysRevD.97.076002,PhysRevA.98.023817} laser beams. For example, tight-focusing can be performed using a parabolic mirror (PM) configuration  \cite{Fillion,PhysRevLett.107.073602,Vallieres:23}, where initially co-propagating laser pulses are focused by reflection into a micrometric focal spot where WMP can take place. Owing to the PM geometry, the angles between propagation directions of laser pulses coming into the focal point can vary from 0 (co-propagation) to $\pi$ (counter-propagation). Another type of geometry is when a strong pump pulse and a weaker probe initially propagate in opposite directions \cite{Lund, Roso_2022}. In this case, the pulses must be very carefully aligned and synchronized, while the detector should be placed behind the pump laser to minimize the signal from pump photons. In all these cases, the intensity requirements to maximize the photon signal will make for a small interaction region, on the order of a few laser wavelengths.

Now we recall that this interaction occurs in a rarefied ionized gas. Therefore, the scattering of photons either by electrons or by the QED-vacuum will both be present and one of them will predominate for some gas pressure and laser intensity. To model these processes, Maxwell's equations need to be solved in combination with either the kinetic (or dynamic) equations for gas or QED expressions for vacuum polarization and magnetization. This will be discussed in details in the next sections. It is challenging to simulate exactly the 3D-geometry of experimental configurations mentioned above, as this would make the computational model too memory-heavy and slow. Instead, we make some simplifications in our model to reflect the key features of the process and its proper scaling, see Figure~\ref{model} for an illustration:
\begin{figure}[b]
	\centering{
		\includegraphics[scale=0.58,trim={10 0 0 10},clip]{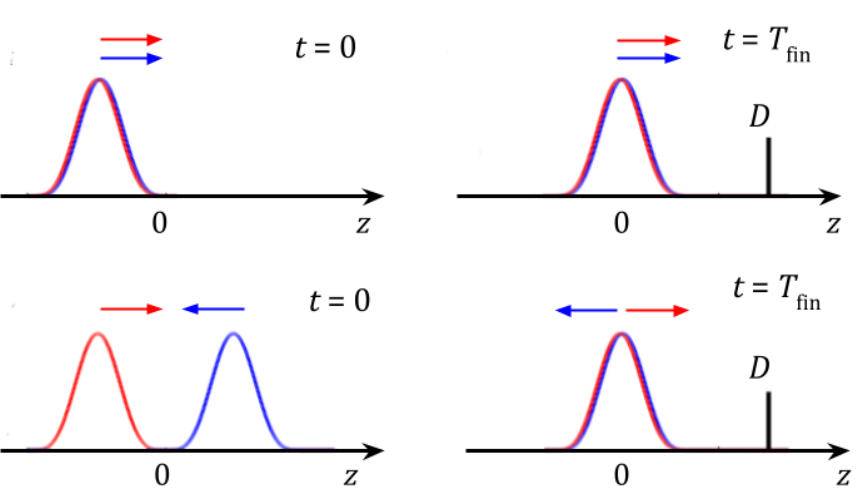}}
	\caption{Interaction models for co-propagating (top row) and counter-propagating (bottom row) incident laser pulses, shown schematically through their intensity profiles. Colored arrows indicate the direction of propagation of pulses shown in the corresponding color. The interaction begins at $t=0$ when the envelopes approach the focal spot ($z=0$ in the picture) and occurs before the time $T_\mathrm{fin}$ when both maxima pass through the focus. Further, the free propagation of the generated photons to the detector ``\textit{D}'' is considered. For case 
	(A), the red pulse corresponds to 800 nm, and blue to 400 nm; for 
	(B), both pulses have identical properties.} \label{model}
\end{figure}
\begin{enumerate}
	\item We consider only the interaction of two short pulses along one axis (the $z$-axis): co-propagating or counter-propagating. 
	\item We assume that interactions only take place in the immediate vicinity of the focal spot. The scattered field further propagates towards the detector without interacting with incident pulses.
	\item We choose the pulse width envelope to be comparable to the size of the interaction region: the full width at half maximum (FWHM) parameter for the intensity profile is $\tau_\mathrm{FWHM}=25$ fs or 7.5 $\mu$m in space, which is a typical value for facilities generating laser ultra-high intensity pulses;
	\item We consider two types of incident pulses:
	\begin{enumerate}
		\item[(A)]\label{ita} The center frequencies of the interacting envelopes are noticeably different: $\lambda_1=800$~nm (typical central value for a high intensity source) and $\lambda_2=400$~nm that can be obtained by doubling of $\lambda_1$ \cite{Wang:17}. The spectra of both pulses have clear maxima at the corresponding $\omega_1$ and $\omega_2$ with a fast decay on the sides.
		\item[(B)]\label{itb} Both pulses contain identical sets of wavelengths in the range $\lambda~\in~[720, 880]$~nm, so their spectra have flat tops centered at 800~nm. This case can be realized more easily in experiments than 
		(a) because it does not require a second laser beam.
	\end{enumerate}
\end{enumerate}

For us, the most interesting control parameters of these models are the peak laser intensity, and the residual pressure in the vacuum chamber before the laser pulse arrives there. 

We now provide mathematical expressions for electromagnetic components of all the incident pulses mentioned above. 
For case (A), we consider the interaction of two linearly polarized electromagnetic waves with an envelope: $(E^{\mathrm{laser}}_{ix}(z,t), B^{\mathrm{laser}}_{iy}(z,t))$, $i=1,2$, propagating in vacuum along the $z$-axis with dispersion relation $\omega_i=k_ic$. Here, we use notations $E_{0i}$ or $B_{0i}$ for amplitudes, which have different units for models in vacuum and gas (see Sections~\ref{plasma} and \ref{vacuum} for details). 
We assume that one laser pulse marked with the number ``1'' always propagates in the positive direction of the $z$-axis (in Fig.~\ref{model} it is shown in blue), while pulse ``2'' is either co-propagating with the first one or counter-propagating (red profiles in Fig.~\ref{model}). Thus, for case 
(A), the incident wave is given by
\begin{eqnarray}\label{E_one_mono}
\nonumber &&E_{ix}^{\mathrm{laser}}(z,t) =  E_{0i}\sin(k_i(z-z_{0i})\mp\omega_it)\times\\
\nonumber &&\:\left\{\begin{array}{c} 
\sin^2\Big(\pi\dfrac{z-z_{0i}\mp ct}{c\sigma}\Big), z_{0i}\pm ct+\sigma>z>z_{0i}\pm ct\\
0, \quad \text{otherwise} \end{array} \right.\\
\end{eqnarray}
where we applied a trigonometric envelope \cite{Barth} with parameter $\sigma$ which can be expressed in terms of the FWHM parameter for the pulse intensity as 
\begin{equation*}
\sigma = \tau_{\text{FWHM}}/\Big(1-\dfrac{2}{\pi}\arcsin\Big(\dfrac{1}{\sqrt[4]{2}}\Big)\Big).
\end{equation*} 
In particular, a typical pulse duration of $\tau_{\text{FWHM}}$ = 25~fs, as mentioned above, corresponds to $\sigma\approx69$ fs. Parameter $z_{0i}$ sets the position of the left edge of each envelope at $t=0$. Regarding the choice of sign in the arguments of $\sin$ and $\sin^2$ in \eqref{E_one_mono}, for pulse ``1'' it is always ``$-$'', while for pulse ``2'' we must choose ``$-$'' in case of co-propagation and ``$+$'' for counter-propagation.

\begin{figure}[t]
	\centering{
		\includegraphics[scale=0.52,trim={4 0 0 0},clip]{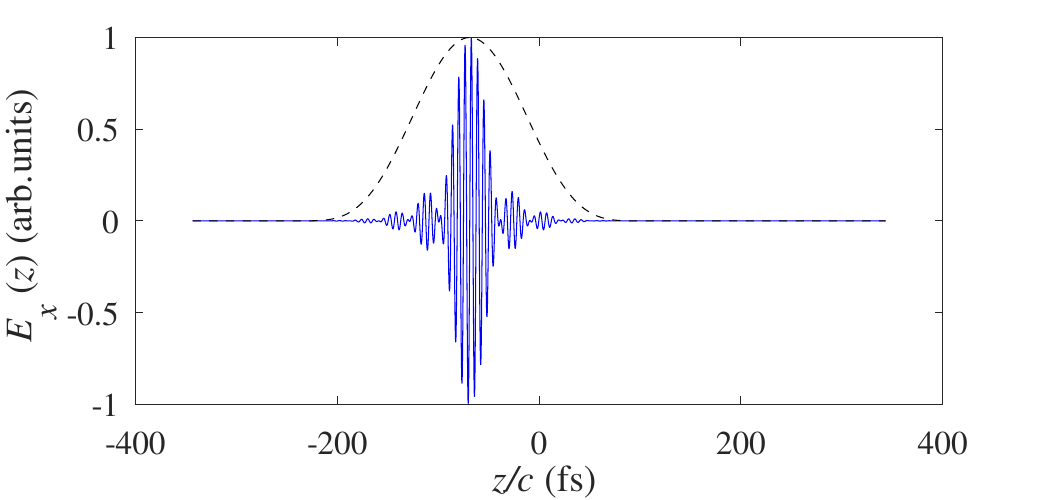}}
	\caption{The electromagnetic field for broadband pulse \eqref{E_one_band} at $t=0$, the dashed line shows the field envelope profile.} \label{init_brd}
\end{figure}

\begin{figure}[t]
	\centering{
		\includegraphics[scale=0.52,trim={5 0 0 0},clip]{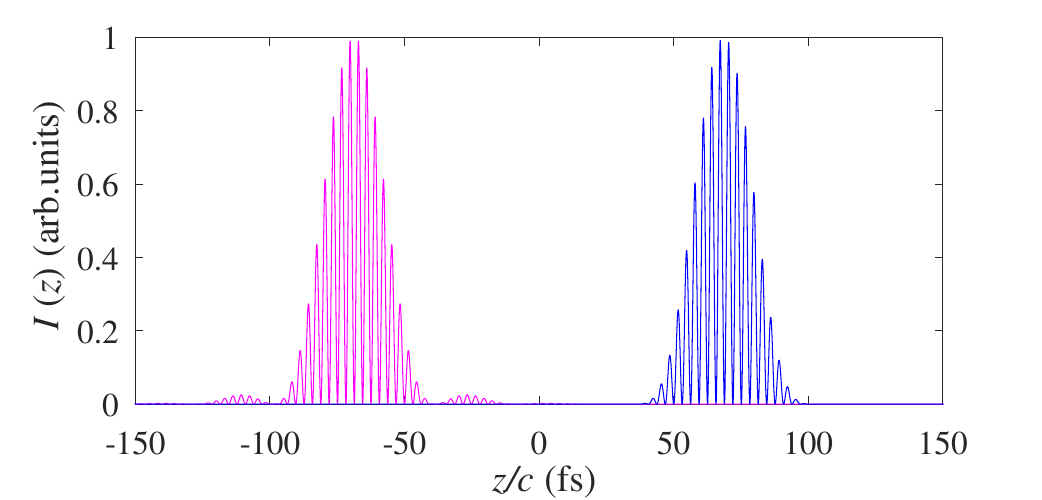}}
	\caption{Incident pulse intensity profiles at $t=0$, defined as $I(z)=E_x^2(z)$, are shown for case (A), according to Eq~\eqref{E_one_mono} with $\lambda_1 = 800$ nm, $\sigma = 68.7$ fs - on the right in blue, and case~(B), Eq.~\eqref{E_one_band}, where $\lambda\in[720 , 880 ]$ nm, $\sigma = 68.7$ fs  - on the left in magenta.} \label{init_intens}
\end{figure}

\begin{figure}[h!]
	\centering{
		\includegraphics[scale=0.52,trim={5 0 0 0},clip]{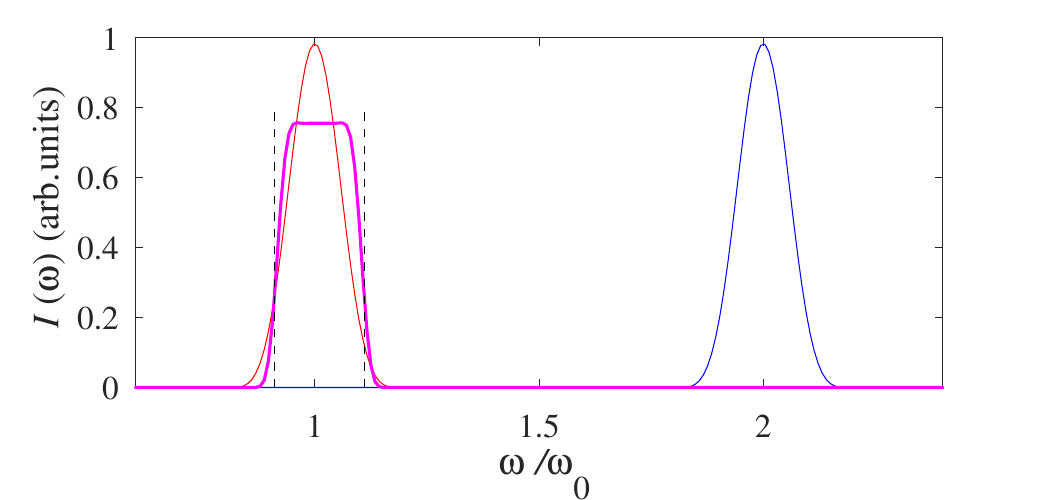}}
	\caption{Spectral intensities of incident laser pulses corresponding to the electric field \eqref{E_one_mono} - thin lines: on the left (red) for $\omega_1=2\pi c/\lambda_1$, on the right (blue) for $\omega_2=2\pi c / \lambda_2$; and electric field \eqref{E_one_band} - thick line (magenta). The dotted lines show the region $[\omega_{\mathrm{min}}, \omega_{\mathrm{max}}]$ for the broadband train. All frequencies are normalized to $\omega_0=2\pi c /\lambda_1$.} \label{init_spc}
\end{figure}

For case (B) which consists of 
a broadband train with wavelengths in between $\lambda_{\mathrm{min}}=720$ nm and $\lambda_{\mathrm{max}} = 880$ nm, hence $\omega_{\mathrm{min}}=2\pi c/\lambda_{\mathrm{max}}$, $\omega_{\mathrm{max}}=2\pi c/\lambda_{\mathrm{min}}$ and central frequency $\omega_c = \pi c  (\lambda_{\mathrm{min}}~+~\lambda_{\mathrm{max}})/\lambda_{\mathrm{min}}\lambda_{\mathrm{max}}$, we assume that the envelope is non-zero for $z_{0i}\pm ct+5\sigma>z>z_{0i}\pm ct$. Then the field components are given by
\begin{eqnarray}\label{E_one_band}
\nonumber &&E_{ix}^{\mathrm{laser}}(z,t) = E_{0i}\sin(k_c(z-z_{0i})\mp\omega_ct)\times\\
\nonumber&&\qquad\dfrac{\sin\Big[(N+1/2)(\Delta k (z-z_{Mi})\mp\Delta\omega t) \Big]}{\sin\big[(\Delta k (z-z_{Mi})\mp\Delta\omega t)/2\big]}\times\\ 
&&\qquad\qquad\sin^4\Big(\pi\dfrac{z-z_{0i}\mp ct}{5c\sigma}\Big),
\end{eqnarray}
where $k_c=\omega_c/c$, $\Delta k = (k_{\mathrm{max}}-k_{\mathrm{min}})/2N$, $\Delta \omega = (\omega_{\mathrm{max}}-\omega_{\mathrm{min}})/2N$. In simulations, we choose $N=15$ which corresponds to $2N+1=31$ spectrum components very close to each other. In Eq. \eqref{E_one_band}, $z_{0i}$ stands again for the left edge position of the envelope, and $z_{Mi}= z_{0i} + 5c\sigma/2$ is the position of its maximum. The support of the $\sin^4$-envelope is five times wider than for (A)
to localize the pulse in space/time without loosing frequencies, as seen in Fig.~\ref{init_brd}. The pulse duration at FWHM is determined by the bandwidth, i.e. the second factor in the expression \eqref{E_one_band}. Thus, for the chosen parameters, we have the same FWHM values ($\approx$ 25 fs) for the intensities in both (A) and (B), see Fig.~\ref{init_intens}. In addition, Figure~\ref{init_spc} shows the spectral intensities of the original envelopes, defined as $I(\omega) = |\hat{E}_x(\omega)|^2$.


The corresponding magnetic field, for propagation in the positive direction $z$, is given by (in the SI system):
\begin{equation}
B_{iy}^{\mathrm{laser}}(z,t) = E_{ix}^{\mathrm{laser}}(z,t)/c,\quad i = 1,2
\end{equation}
and for propagation in the negative direction of the $z$-axis:
\begin{equation}
B_{2y}^{\mathrm{laser}}(z,t) = -E_{2x}^{\mathrm{laser}}(z,t)/c,
\end{equation}
given that in the incident pulses $B_{i0}=E_{0i}/c$.

\section{System of Vlasov-Maxwell equations for rarefied gas}\label{plasma}
Equations and parameters in this section use Gaussian units (aka CGS) unless otherwise specified. We consider two laser pulses of ultra-high intensity (more than $10^{19}$ W/cm$^2$) propagating in extremely rarefied helium gas ($P<10^{-10}$ Torr) along the $z$-axis. Before turning on the laser, we can calculate the spatially homogeneous number density $n_\mathrm{He}$, knowing the initial pressure $P$ and considering room temperature of $T_0=293$ K:
\begin{equation}
n_\mathrm{He}(\mathrm{cm}^{-3})\approx 3.3\cdot10^{6}\cdot\dfrac{P(\mathrm{Torr})}{10^{-10}}.
\end{equation}
Our model does not include considerations of the entire ionization process, but it assumes that at intensities where WMP in vacuum is noticeable, the gas is quickly ionized, so that the initial spatially homogeneous number densities are $n_{\mathrm{He}^{2+}0}=n_\mathrm{He}$ and $n_{\mathrm{e}0} = 2n_\mathrm{He}$. 
As we discuss in Appendix~\ref{plasma_alg}, the domain we consider in our simulations is an order of magnitude larger than the interaction region; it includes both envelopes of interacting pulses, and also unperturbed space where the pulses propagate freely.  The length of the domain is set to $L_z^\mathrm{max}\approx 200$ $\mu$m. Thus, we can estimate the number of electrons in a sphere of radius $L_z^\mathrm{max}/2$ as
\begin{equation}
	N_\mathrm{e}\approx 28\cdot\dfrac{P(\mathrm{Torr})}{10^{-10}},
\end{equation}
which may be too small to apply kinetic equations directly as to obtain the latter, we must average over phase spaces containing a large number of particles. To verify this more quantitatively, we now estimate the Debye length and plasma parameter of our system.

At very low densities and rapid ionization we cannot expect the particle momentum distribution to become Maxwellian immediately after ionization. Nevertheless, we can still estimate an effective temperature $T_{s}$, which is equivalent to the average kinetic energy acquired by particles with rest mass $m_s$ and charge $q_s$ in a strong laser field with intensity $I_0$. It is given by $\mathcal{E}_s=m_sc^2(\sqrt{1+2U_{ps}/m_sc^2}-1)$, where $U_{ps}=I_0q_s^2\lambda^2/(2\pi m_sc^3)$ is the ponderomotive energy  for particles of type $s$. From here, we have
\begin{eqnarray}\label{temperature}
\nonumber &&T_s(\text{MeV}) \approx m_sc^2(\text{MeV})\times\\ &&\quad\Big\{\sqrt{1+\dfrac{I_0(\text{W/cm}^2)\lambda(\mu\text{m})^2}{2.74\cdot10^{18}}\Big(\dfrac{q_s}{e}\dfrac{m_\mathrm{e}}{m_s}\Big)^2} -1 \Big\},
\end{eqnarray}
which leads to $T_\mathrm{e}\approx 2$ MeV at $\lambda=800$ nm and $I_0=10^{20}$~W/cm$^2$ (assuming here pre-ionization by the pedestal of the laser pulse). Having estimated the temperature $T_\mathrm{e}$ and density $n_{\mathrm{e}0}$, we obtain a value for the Debye radius that is quite large relative to the domain size $L_z^\mathrm{max}$:
\begin{equation}
\lambda_\mathrm{D}\approx 2.9\cdot 10^2\sqrt{\dfrac{T_\mathrm{e}(\mathrm{MeV})}{P(\mathrm{Torr})/10^{-10}}}.
\end{equation}
Moreover, in a non-equilibrium plasma, this screening parameter can be even larger than the one estimated through the equilibrium temperature \cite{Fahr}. Therefore, a fluid/continuum description cannot be applied to our system since we cannot expect quasi-neutrality on the scales of our interaction region. Then, we can consider the plasma parameter, given by $\Lambda \sim n_{e} \lambda_\mathrm{D}^{3}$:
\begin{align}
    & \Lambda\approx 1.61 \cdot 10^{14} \cdot  T^{\frac{3}{2}}_\mathrm{e}(\mathrm{MeV}) \left(\dfrac{P(\mathrm{Torr})}{10^{-10}} \right)^{-\frac{1}{2}}.
\end{align}
For all the configurations considered, even at the lowest gas densities, we have that $\Lambda \gg 1$, which is the condition for a weakly coupled plasma. The regime where $\Lambda \gg 1$ and $\lambda_{D} \gg L$ corresponds to a dilute gas, for which the dynamics are essentially a collection of driven particles with weak self-interaction.

Therefore, describing the system mathematically requires free particles $s = \{$e$^{-}$, He$^{2+}\}$ with dynamic coordinate ${\bf r}_s$ and momentum ${ \bf p}_s$ changing under the action of the relativistic Lorentz force
\begin{align}
\frac{d\mathbf{r}_s}{dt} &= \frac{\mathbf{p}_s}{m_s \gamma_s}, \\
\frac{d\mathbf{p}_s}{dt} &= q_s\left[\mathbf{E} + \frac{\mathbf{p}_s}{m_s \gamma_s} \times \mathbf{B}\right],
\end{align}
where ${\bf E}$, ${\bf B}$ is the electromagnetic field, $q_s$ and $m_s$, as above, are the charge and mass of the particle $s$, $\gamma~=~\sqrt{1 + {\bf p}_s^{2}/m_s^{2}c^{2}}$. These equations must be supplemented by initial conditions. For particles in a gas, the initial conditions are random and characterized by an initial distribution $F_{0}({\bf r}_s, {\bf p}_s)$ which gives the probability to find a particle in phase space at the initial time. Such system of ordinary differential equation with random initial conditions can be converted to a partial differential equation using Liouville's theorem \cite{soong1973random}. Through this procedure and specializing to 1D (assume homogeneity in $x$ and $y$), one recovers the Vlasov equation
\begin{equation}\label{vlasF}
\partial_tF_s+\dfrac{p_{sz}}{m_s\gamma_s}\partial_zF_s+q_s\Big({\bf E}+\dfrac{{\bf p}_s\times {\bf B}}{m_s\gamma_sc}\Big)\cdot\nabla_{\bf p} F_s = 0,
\end{equation}
where we introduced the distribution function $F_s(z,{\bf p},t)$ for particles of species $s$. In contrast to a plasma, where $\lambda_{D} \ll L$ and where the larger number of particles allows for a coarse-grained description of the dynamics with a single realization, this statistical description is only valid for a large number of realizations $N_{R} \gg 1$. In this formalism, the Vlasov equation describes the evolution of the ensemble average, not the exact dynamics of any single system. Individual realizations are dominated by strong particle-particle fluctuations, but after averaging over many independent realizations, the fluctuations cancel, and the ensemble-averaged field is smooth and predictive.

In 1D, we restrict electromagnetic field components to ${\bf E}=(E_x, 0, E_z)$ and ${\bf B} = (0, B_y, 0)$. Then the evolution of the electromagnetic field due to scattering by the particles of the medium is described by the Maxwell equations as follows:
\begin{eqnarray}
\label{Ampere}
\partial_t E_x(z,t) &=& -c\partial_z B_y(z,t)-4\pi \langle J_x(z,t) \rangle,\\
\label{Faraday}
\partial_t B_y(z,t) &=& -c\partial_z E_x(z,t),\\
\label{Ampere0}
\partial_t E_z(z,t) &=& -4\pi \langle J_z(z,t) \rangle,
\end{eqnarray}
where the current densities $\langle J_x \rangle$ and $\langle J_z \rangle$ are ensemble averaged and must be defined via medium properties characterized by the distribution functions. 

Under the given configuration of electromagnetic fields, an exact factorization of the solution for \eqref{vlasF} is \cite{Ghizzo90, Huot, Strozzi2004}
\begin{equation}\label{separation}
F_s(z,{\bf p},t) = \delta(P_{sx})f_s(z,p_z,t),
\end{equation}
where the reduced 1D-1V distribution function $f_s$ obeys another Vlasov equation  
\begin{equation}\label{vlasov}
\partial_tf_s+\dfrac{p_z}{m_s\gamma_s}\partial_zf_s+q_s\Big(E_z+\dfrac{p_{x}B_y}{m_s\gamma_sc}\Big)\partial_{p_z}f_s=0,
\end{equation}
where the canonical momentum is $P_{sx}=p_{sx}+\dfrac{q_s}{c}A_x$. For the momentum component transverse to the propagation of the wave, we write (further omitting $s$ in the momentum index)
\begin{equation}\label{Px}
p_x(z,t) = -\dfrac{q_s}{c}A_x(z,t).
\end{equation}
Taking the partial derivatives of \eqref{Px} with respect to $z$: 
\begin{equation}\label{tranPx}
\partial_z p_x(z,t) = -\dfrac{q_s}{c}B_y(z,t).
\end{equation}
Now for the current densities in equations, \eqref{Ampere0} and \eqref{Ampere} we have:
\begin{align}
\label{current}
\nonumber &\langle J_x(z,t) \rangle = \sum_s q_s \iint_{-\infty}^{\infty}\dfrac{p_x}{\gamma_sm_s}F_s(z,{\bf p},t)dp_xdp_z=\\
\nonumber&\:=\sum_s \dfrac{q_sp_x(z,t)}{m_s} \int_{-\infty}^{\infty}\dfrac{f_s(z,p_z,t)}{\sqrt{1+(p_x^2(z,t)+p_z^2)/m_s^2c^2}}dp_z,\\
\\
\label{current_z}
\nonumber &\langle J_z(z,t) \rangle = \sum_s \dfrac{q_s}{m_s} \int_{-\infty}^{\infty}\dfrac{p_zf_s(z,p_z,t)}{\sqrt{1+(p_x^2(z,t)+p_z^2)/m_s^2c^2}}dp_z.\\
\end{align}
Thus the Vlasov-Maxwell system of equations \eqref{Ampere0}-\eqref{Faraday}, \eqref{vlasov}, \eqref{tranPx}-\eqref{current_z} becomes self-consistent \cite{lifshitz1995physical}, and we only need to add the relevant initial and boundary conditions. 

Looking ahead,  we verified during the simulations that final results, such as radiation spectra, do not depend on the $T_s$ of the initial electron distribution (actually even on its shape), if it is taken to be equal to or less than 1~MeV. In fact, particle acceleration with strong laser fields yields much higher kinetic energies  than the thermal initial energy distribution, making it of little importance, as long as it is much smaller than the final particle energy. That is why in simulations, we simply use Maxwellian distributions (in $p_z$) with temperatures $T_\mathrm{e}=1$~MeV and $T_\mathrm{He}=100$~keV as an initial condition. We also assume that within our computational domain, all fields $E_x, B_y, E_z$ and momentum $p_x$ are equal to zero before the laser is turned on. The initial envelope of the incident laser field is described by \eqref{E_one_mono} or \eqref{E_one_band} at $t=0$.

Although we consider the high intensity range, the electron dynamics equation \eqref{vlasF} does not include radiation-reaction forces. Following Refs. \cite{PhysRevLett.105.220403, RevModPhys.84.1177, Blackburn}, we can estimate the importance of radiation losses using the classical radiation reaction parameter:
\begin{align}
	R_c = 0.13\lambda(\mu\mathrm{m})\Big(\frac{\mathcal{E}_\mathrm{e}(\mathrm{MeV})}{500}\Big)\Big( \frac{I_0(\mathrm{W/cm}^2)}{10^{22}}\Big),
\end{align}
where $\mathcal{E}_\mathrm{e}$ is the energy of the electron. Since the initial kinetic energy of electrons is 1 MeV in our simulations while $\lambda=800$ nm and the maximum intensity is $I_0=10^{23}$ W/cm$^2$, we get $ R_c\approx0.003$. This is less than the threshold value of 0.024 at which radiation-reaction must be included in the model, so it can be neglected. As the electron energy increases as they are accelerated in the laser pulses, the radiation losses can become increasingly significant for the most energetic of them. In our calculations we do not fall into the radiation-dominated regime, where $R_c > 1$. 

In addition to the classical parameter $R_c$, a quantum parameter proportional to $\hbar$ is used to measure the influence of QED effects:
\begin{align}
	\chi = 0.29\Big(\frac{\mathcal{E}_\mathrm{e}(\mathrm{MeV})}{500}\Big)\Big( \frac{I_0(\mathrm{W/cm}^2)} {10^{22}}\Big)^{1/2}.
\end{align}
This parameter is responsible for the amount of electron energy that a single photon can carry away. If $\chi \geq 1$, then the fraction of electron energy losses is significant. In our simulation, at the maximum considered intensity $10^{23}$~W/cm$^2$ and the maximum achievable electron energies, see Fig.~\ref{distr} in Appendix~\ref{plasma_alg}, this parameter is less than 0.9. So, the quantum effects can play important role but for statistically rare particles. We return to the possible influence of radiation reaction effects on the results for ionized gas in the Conclusion section.

Our computational model uses the  scattered field formalism \cite{Kunz}, which divides the fields into the sum of the ``laser'' field analytically calculated from \eqref{E_one_mono} or \eqref{E_one_band} and the generated or ``scattered'' field, which is calculated numerically:
\begin{align}
\label{e_tot}
E_x(z,t)&=E_{1x}^{\mathrm{laser}}(z,t)+E_{2x}^{\mathrm{laser}}(z,t)+E_x^{\mathrm{scatt}}(z,t),\\
\label{b_tot}
B_y(z,t)&=B_{1y}^{\mathrm{laser}}(z,t)+B_{2y}^{\mathrm{laser}}(z,t)+B_y^{\mathrm{scatt}}(z,t).
\end{align}
Because the laser fields are analytical solutions to the free Maxwell equation, they vanish from Eqs. \eqref{Ampere}, \eqref{Faraday} and \eqref{Ampere0}. 
Also we introduce transfer fields as
\begin{equation}
E^\pm(z,t) = E_x(z,t)\pm B_y(z,t),
\end{equation}
from where, obviously, one can extract the scattered part $E^\pm_{\mathrm{scatt}}(z,t)$. For the convenience of numerical calculations, we normalize the model quantities to their characteristic values, while retaining the same notation, see Table~\ref{plasma_inits}, where we introduced the critical number density $n_\mathrm{cr}=m_e\omega_0^2/4\pi e^2\approx1.74\cdot10^{21}$ cm$^{-3}$ starting from which electromagnetic waves cannot propagate in plasma \cite{Chen}. 
\begin{table}[b]
	\begin{center}
		\begin{tabular}{ccc}\hline \hline
			variable & model units (CGS) &  value in SI units  \\	\hline
			$t$ & $1/\omega_0=\lambda_1/2\pi c$ & $4.25\cdot10^{-16}$ s\\						
			$z$ & $c/\omega_0=\lambda_1/2\pi$  &  $1.27\cdot10^{-7}$ m \\												
			$E_x$, $E_z$ & $m_\mathrm{e}c\omega_0/e$ & $4.01\cdot10^{12}$ V/m \\						
			$B_y$ & $m_\mathrm{e}c\omega_0/e$ & $1.34\cdot10^4$ T \\					
			$p_z$ & $m_\mathrm{e}c$ & $2.73\cdot10^{-22}$ kg$\cdot$m$/$s \\
			$f_s$ & $n_\mathrm{cr}/m_\mathrm{e}c$ & $6.37\cdot10^{48}$ s$/$(kg$\cdot$m$^4$) \\
			$\langle J_x \rangle, \langle J_z \rangle$ & $en_\mathrm{cr}c$ & $8.36\cdot10^{16}$ A/m$^2$ \\
			\hline \hline																										
		\end{tabular}
		\caption{Normalization units for the system of Vlasov-Maxwell equations. \label{plasma_inits}}
	\end{center}
\end{table}
Thus, we present the normalized model for self-consistent fields:
\begin{align}
\label{tranE2}
&(\partial_t\pm\partial_z) E^\pm_{\mathrm{scatt}}(z,t) = -\langle J_x(z,t) \rangle,\\
\label{Ampere2}
&\partial_t E_z(z,t) = - \langle J_z(z,t) \rangle ,\\
\label{vlasov2}
\nonumber&\partial_tf_s+\dfrac{p_z}{M_s\gamma_s(p_x,p_z,M_s)}\partial_zf_s+\\
&\qquad Q_s\Big(E_z+\dfrac{p_x}{M_s\gamma_s(p_x,p_z,M_s)}B_y\Big)\partial_{p_z}f_s=0,\\
\label{px_evol_norm}
&\partial_zp_x(z,t) = -Q_sB_y(z,t),\\
\label{current2}
\nonumber&\langle J_x(z,t) \rangle= \sum_s \dfrac{Q_sp_x(z,t)}{M_s}\! \int_{-\infty}^{\infty}\!\dfrac{f_s(z,p_z,t)}{\gamma_s(p_x(z,t),p_z,M_s)}dp_z,\\
\\
\label{current2_z}
&\langle J_z(z,t) \rangle = \sum_s \dfrac{Q_s}{M_s} \int_{-\infty}^{\infty}\dfrac{f_s(z,p_z,t)p_z}{\gamma_s(p_x(z,t),p_z,M_s)}dp_z,
\end{align}
where, for brevity of notation in the equation~\eqref{vlasov2}, we omitted the argument $(z,p_z,t)$ of the distribution function and the argument $(z,t)$ of the fields and the $x$-component of the particle momentum. In the system we explicitly introduced normalized masses $M_s = m_s/m_\mathrm{e}$ and charges $Q_s=q_s/e$, so that the relativistic factor becomes
\begin{equation*}
\gamma_s(p_x(z,t),p_z,M_s)~=~\sqrt{1+(p_x^2(z,t)+p_z^2)/M_s^2}.
\end{equation*}

The numerical algorithm for solving this system of equation is given in Appendix~\ref{plasma_alg}. For seven decades, one of the most popular methods for solving Vlasov-Maxwell systems has been the particle-in-cell (PIC) method \cite{Arber_2015}, mainly because of its simplicity and speed. Its main disadvantage is also well-known: a high level of numerical noise, especially in cases where the particle density is very low. That is why various Euler or semi-Lagrangian methods for solving the Vlasov-Maxwell equations, slower but more accurate, have also become popular in the last three decades, see e.g. \cite{Ghizzo90, Durran, Grizzo03, Huot, Strozzi2004, Palmroth2018, fletcher2019semi}. For our purpose of modeling WMP in a rarefied gas, the semi-Lagrangian approach is more suitable. However, due to performance and memory requirements, we still have to restrict our model to 1D-1V geometry.

To conclude this section, we discuss observable values. To study the result of the interaction, we calculate the spectral intensity of the generated photons $I^{\mathrm{scatt}}(\omega) = |\hat{E}^{\mathrm{scatt}}_x(z_D,\omega)|^2$, where $\hat{E}^{\mathrm{scatt}}_x(z_D,\omega)=\mathcal{F}\big\{E^{\mathrm{scatt}}_x(z_D,t)\big\}$ is the Fourier transform of the scattered electric field at position $z_D$ of the detector, see Figure~\ref{model}. Another important parameter is the photon number density generated per surface element $d\mathcal{S}$ during the time of interaction,  $T_\mathrm{fin}$ in Figure~\ref{model}. This can be estimated using the Poynting theorem:
\begin{equation}\label{Poynting}
\dfrac{d\mathcal{N}}{d\mathcal{S}}=\dfrac{T_\mathrm{fin}c}{2\pi}\!\sum_{n=N_{\mathrm{min}}}^{N_{\mathrm{max}}}\!\dfrac{\text{Re}\big[\hat{E}^{\mathrm{scatt}}_x(z_D,\omega_n)\hat{B}^{\mathrm{scatt}}_y(z_D,\omega_n) \big]}{\hbar\omega_n},
\end{equation}
where the summation is over all frequencies of the generated photons $[\omega_{\mathrm{min}}, \omega_{\mathrm{max}}]$.

\section{Modified Maxwell equations in the vacuum}\label{vacuum}
Here we consider the model of photon generation in vacuum. Throughout this Section we use natural units: $\hbar = c = \varepsilon_0 = \mu_0 = m_\mathrm{e} = 1$, $e = \sqrt{4\pi \alpha}$, where $\alpha$ is the fine structure constant. Units of length, time, and critical Schwinger fields at which Maxwell's equations become strongly nonlinear \cite{Schwinger} are given in Table~\ref{vacuum_units}
\begin{table}[b]
	\begin{center}
		\begin{tabular}{ccc}\hline \hline
			variable & model units (natural) &  value in SI units  \\	\hline
			$t$ & $\tau_C = \hbar/m_\mathrm{e}c^2$ & $1.288\cdot 10^{-21}$ s\\						
			$z$ & $\lambda_C = \hbar/m_\mathrm{e}c$  &  $3.862\cdot 10^{-13}$ m \\												
			$E_x$, $E_z$ & $E_{S}=m_\mathrm{e}^2c^3/e\hbar$ & $1.323\cdot 10^{18}$ V/m \\						
			$B_y$ & $B_{S}=m_\mathrm{e}^2c^2/e\hbar$ & $4.414\cdot 10^{9}$ T \\								
			\hline \hline																										
		\end{tabular}
		\caption{Units for modified Maxwell equations for vacuum. \label{vacuum_units}}
	\end{center}
\end{table}

Vacuum polarization in QED is understood as the formation of virtual particle-antiparticle pairs under the action of a strong enough electromagnetic field. Photon-photon interactions occur because of the creation of these pairs, which, in turn, affects the particle-antiparticle dipoles and make the process nonlinear. Since the laser fields considered are 3-4 orders of magnitude lower than the Schwinger limit, we do not need to use the full Euler-Heisenberg action to take QED effects into account, it is enough to consider their lowest order radiative corrections \cite{berestetskii1982quantum}. These corrections were obtained for the first time by Heisenberg and Euler under the conditions that variations of the electromagnetic fields are negligible over the Compton spatial $\lambda_C$ and time $\tau_C$ scales  \cite{Heisenberg,dunne2005heisenberg}.
In this case, Maxwell's equations in vacuum, which include vacuum polarization effects, are given by \cite{Mckenna, Moulin99, Domenech}:
\begin{equation}\label{Max_base}
\left\{\begin{array}{ccc}
\partial_t{\bf B}&=&-\nabla\times {\bf E}\\
\partial_t({\bf E}+ {\bf P})&=&\nabla\times ({\bf B}-{\bf M}),
\end{array}\right.
\end{equation}
where the nonlinear macroscopic polarization and magnetization are calculated through the Heisenberg-Euler Lagrangian
\begin{equation}\label{pol_mag}
{\bf P} = \nabla_{\bf E}\mathcal{L}_{HE},\qquad {\bf M} = \nabla_{\bf B}\mathcal{L}_{HE}.
\end{equation}
Although in general this Lagrangian includes an integral, in the weak-field expansion ($E \ll 1$) with the help of the Taylor series one arrives at (see e.g. \cite{Domenech})
\begin{eqnarray}\label{weak_field}
\nonumber \mathcal{L}_{HE} &\approx& a_1(4\mathcal{F}^2+7\mathcal{G}^2)\\
\nonumber&+&a_2(8\mathcal{F}^3+13\mathcal{F}\mathcal{G}^2)\\
&+&a_3(48\mathcal{F}^4+88\mathcal{F}^2\mathcal{G}^2+19\mathcal{G}^4),
\end{eqnarray}
where the coefficients in natural units are
\begin{equation}
\quad a_1 = \dfrac{1}{360\pi^2}, \quad a_2 = \dfrac{1}{630\pi^2}, \quad a_3 = \dfrac{1}{945\pi^2},
\end{equation}
while $\mathcal{F}$ and $\mathcal{G}$ are expressed through the Lorentz invariants: 
\begin{eqnarray}\label{Lor1}
\mathcal{F} &=& -\dfrac{(E^2 - B^2)}{2E^2_S} = -4\pi\alpha\dfrac{(E^2 - B^2)}{2},\\
\label{Lor2}
\quad \mathcal{G} &=& \dfrac{{\bf E}\cdot{\bf B}}{E^2_S}=4\pi\alpha\:{\bf E}\cdot{\bf B}.
\end{eqnarray}
Before proceeding, we make two comments on expression \eqref{weak_field}. First, since the Heisenberg-Euler Lagrangian depends on the field through \eqref{Lor1} and \eqref{Lor2}, it means that for any superposition of plane waves copropagating in a vacuum, radiative corrections are zero. Second, the terms in expansion at $a_1$, $a_2$ and $a_3$ are responsible for 4-, 6- and 8-photons contribution, respectively, see e.g. \cite{Domenech, Lindner}. With each new pair of involved photons, the probability that the process takes place decreases in proportion to the coefficient $\alpha$, so one can truncate the expansion as \eqref{weak_field}. It will be verified \textit{a posteriori} in the numerical results of Section~\ref{res} that the highest order contribution (8-photons) is weak.

Back to the Maxwell equation, we assume the same field geometry and field components as in Section~\ref{plasma}, so that ${\bf E}\cdot{\bf B}=0$, and we obtain for ${\bf P} = [P_x(z,t), 0, P_z(z,t)]$:
\begin{eqnarray}\label{PxPz}
\nonumber{\bf P}&=&{\bf E}\dfrac{8\alpha^2}{45}\Big\{(E^2-B_y^2)-\dfrac{24\pi\alpha}{7}(E^2-B_y^2)^2+\\
&&\dfrac{256\pi^2\alpha^2}{7}(E^2-B_y^2)^3\Big\},
\end{eqnarray}
where $E^2 = E_x^2+E_z^2$, and ${\bf M} = [0, M_y(z,t), 0]$:
\begin{eqnarray}\label{My}
\nonumber M_y&=&-\dfrac{8\alpha^2}{45}B_y\Big\{(E^2-B_y^2)-\dfrac{24\pi\alpha}{7}(E^2-B_y^2)^2+\\
&&\dfrac{256\pi^2\alpha^2}{7}(E^2-B_y^2)^3\Big\}.
\end{eqnarray}
Further, from system \eqref{Max_base} we come to the equations
\begin{equation}\label{Max_mod}
\left\{\begin{array}{rcl}
\partial_tB_y+\partial_zE_x &=& 0,\\
\partial_tE_x+\partial_zB_y &=& -J_x,\\
\partial_tE_z &=& -J_z
\end{array}\right.
\end{equation}
where now the ``currents'' are
\begin{eqnarray}
J_x &=& \partial_tP_x-\partial_zM_y,\\
\label{JzPz}
J_z &=& \partial_tP_z.
\end{eqnarray}
Mathematical transformations with \eqref{PxPz}-\eqref{JzPz} lead us to the expressions
\begin{eqnarray}
\label{Jx}
J_x&=&\dfrac{16\alpha^2}{45}\mathcal{R}_1\dfrac{\mathcal{I}_1(1+\delta_z)-\mathcal{C}\mathcal{I}_2}{(1+\delta_x)(1+\delta_z)-\mathcal{C}^2},\\
\label{Jz}
J_z&=&\dfrac{16\alpha^2}{45}\mathcal{R}_1\dfrac{\mathcal{I}_2(1+\delta_x)-\mathcal{C}\mathcal{I}_1}{(1+\delta_x)(1+\delta_z)-\mathcal{C}^2},
\end{eqnarray}
where
\begin{eqnarray}
\nonumber \mathcal{R}_1 &=& 1 -\dfrac{48\pi\alpha}{7}(E^2-B_y^2)+\dfrac{768\pi^2\alpha^2}{7}(E^2-B_y^2)^2\\
\nonumber \mathcal{R}_2 &=& 1 -\dfrac{24\pi\alpha}{7}(E^2-B_y^2)+\dfrac{256\pi^2\alpha^2}{7}(E^2-B_y^2)^2,\\
\nonumber \delta_x &=&\dfrac{8\alpha^2}{45}\big[2\mathcal{R}_1E_x^2 + \mathcal{R}_2(E^2-B_y^2)\big],\\
\nonumber \delta_z&=& \dfrac{8\alpha^2}{45}\big[2\mathcal{R}_1E_z^2 + \mathcal{R}_2(E^2-B_y^2)\big],\\
\nonumber \mathcal{C} &=& \dfrac{16\alpha^2}{45}\mathcal{R}_1E_xE_z,\\
\nonumber \mathcal{I}_1 &=& 2E_xB_y\partial_zE_x-(E_x^2+B_y^2)\partial_zB_y+B_yE_z\partial_zE_z,\\
\nonumber \mathcal{I}_2 &=& E_z(B_y\partial_zE_x-E_x\partial_zB_y).
\end{eqnarray}
In particular, \eqref{Jx}, \eqref{Jz} show that the first term in the expansion for currents has order $\alpha^2$, while the next terms have factors $\alpha^3$ and $\alpha^4$.

Finally, we discuss numerical approaches to solve these equations. One of the most popular method for self-consistent modeling of vacuum polarization is finite-difference time-domain solver, based on the 2nd order generalized Yee scheme. These solvers are multi-dimensional and can be incorporated into PIC model, see e.g. \cite{Grismayer_2021}, \cite{Zhang2025}. Currently, one can find other finite-difference methods up to the 13th order of accuracy \cite{Lindner, Lindner2}. However, for our purposes neither such a high level of approximation nor multidimensionality is required. Instead, we apply the same numerical scheme to solve Maxwell's equation for both gas and vacuum to ensure a proper comparison; see  Appendix~\ref{vacuum_alg} for details. In this model we compute the same observables as described at the end of the Section~\ref{plasma}, although in the natural system of units.

\section{Simulation results}\label{res}
Using the models described in the previous two sections, we calculate and compare the spectral intensities and/or number of photons generated in gas and vacuum by the interaction of two incident laser pulses with the same peak intensities: $I_0=10^{22}$~W/cm$^2$ or $I_0=10^{23 }$~W/cm$^2$. With this purpose, we convert the fields $E_x$ and $B_y$, calculated in both models as normalized quantities, into SI units, see Tables~\ref{plasma_inits} and \ref{vacuum_units}. Before presenting the results of the comparison, we note two important physical findings for further considerations that are confirmed by these numerical models of WMP in vacuum and in gas.

First, according to QED theory, photons are not generated in vacuum if both pulses are co-propagating. This implies that WMP harmonics generated during the co-propagation of two laser pulses in gas can be compared to harmonics generated in vacuum, but with counter-propagation. This also makes sense physically for tightly-focused configurations, where in the focal spot we can simultaneously detect photons generated by the interaction of pulses propagating at different angles. 

Second, our simulations demonstrate that for a fixed intensity of incident pulses $I_0$, a decrease in initial pressure in the vacuum chamber  by an order of magnitude leads to a decrease in the intensity of the scattered field by two orders of magnitude, as follows from equations \eqref{tranE2} and \eqref{current2}. The latter circumstance is convenient for estimating the pressure at which the generation of harmonics in vacuum and gas becomes comparable. 

\subsection{WMP photons from interaction of two laser envelopes with distinct central frequencies}\label{dist}
\begin{figure*}
	\centering{
		\includegraphics[scale=0.9,trim={0 0 0 0},clip]{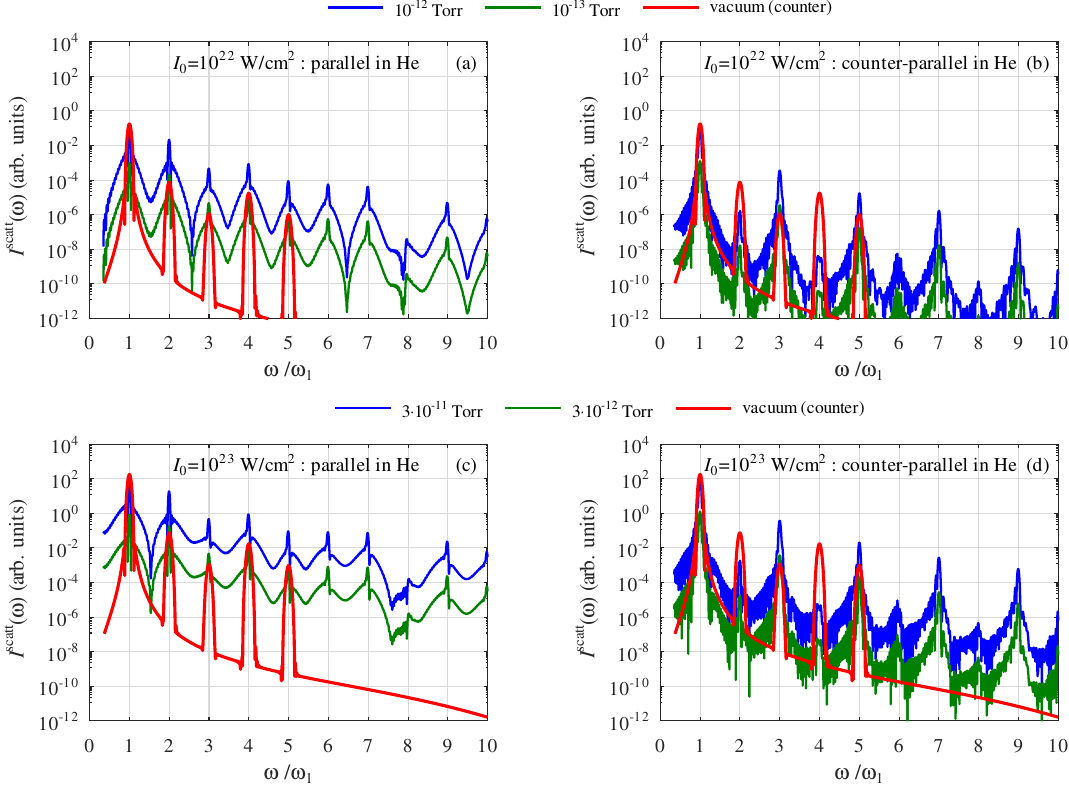}}
	\caption{Spectra of the scattered field generated by the interaction of two envelopes given by formula~\eqref{E_one_mono} with wavelengths $\lambda_1=800$ nm and $\lambda_2=400$ nm and intensities $I_0$ either (a),(b) $10^{22}$ W/cm$^2$ or (c),(d) $10^{23}$ W/cm$^2$. 
	Solutions in helium are shown in inserts (a),(b) for pressures $P=10^{-12}$ Torr - in blue and $P=10^{-13}$ Torr - in green, and in inserts (c),(d) for $P=3\cdot10^{-11}$ Torr - in blue and $P=3\cdot10^{-12}$ Torr - in green; inserts (a),(c) correspond to  co-propagating and (b),(d) to counter-propagating pulses in helium. For comparison, vacuum solutions are shown in red for the same envelopes, the same intensities, but always for counter-propagating pulses. The frequency axis is normalized to $\omega_1=2\pi c/\lambda_1$.} \label{mono_spec}
\end{figure*}

For case of interaction of type (A) envelopes - according to the Section~\ref{num_exp}, given by formula~\eqref{E_one_mono}, with central wavelengths of $\lambda_1= 800$ nm and $\lambda_2 = 400 $ nm, we can observe the  theoretically predicted WMP harmonics in the numerical results, see Figs.~\ref{mono_spec}~(a)-(d) and Table~\ref{fwmf}. 
\begin{table}[b]
	\begin{center}
		\begin{tabular}{ccc}\hline \hline
			notation & formula & value in $\omega_1$ \\	\hline
			$\omega_{1}$ & $2\pi c/ \lambda_1$ & 1\\						
			$\omega_{2}$ & $2\pi c/ \lambda_2$  & 2 \\												
			$\omega_{a}$ & $2\omega_2-\omega_1$ & 3  \\						
			$\omega_{b}$ & $2\omega_a-\omega_2$ & 4 \\					
			$\omega_{c}$ & $2\omega_a-\omega_1$ & 5 \\\hline \hline																										
		\end{tabular}
		\caption{The expected WMP peak positions for interaction of the laser pulses with distinct central frequencies $\omega_{1}$ and $\omega_{2}$. \label{fwmf}}
	\end{center}
\end{table}
Two laser beams of different frequencies interact with currents in fully ionized helium and generate the first sum- and difference-harmonics. These new harmonics interact with the fundamental ones, generating WMP peaks of the following orders.
In gas, both co- and counter-propagating pulses can generate harmonics, although there are differences in shape between them. Thus, the decay of spectral intensity is faster during counter-propagation, due to the fact that the interaction of maximum fields occurs in a shorter time than during co-propagation, when the intensity of new peaks $\omega_{\{a,b, c,\dots\}}$ increases proportionally to the propagation length.
Another feature of the spectra for counter-propagation  in gas is that the fundamental harmonic $\omega_1$ and then $\omega_a$, $\omega_c,\dots$ are much more pronounced than $\omega_2$, $\omega_b,\dots$, see Figure~\ref{mono_spec} (b),(d). This may be a consequence of the fact that in our geometry the fundamental harmonic propagates towards the detector, while the pulse with $\omega_2$ propagates in the opposite direction, see Figure \ref{model}. Whether these features are significant for real measurements depends on the design of the experimental setup.
Note that in Fig.~\ref{mono_spec} the number of generated harmonics in vacuum is always smaller than in gas in any case of propagation. The reason for this is that we consider only three terms in expansion for Heisenberg-Euler Lagrangian \eqref{weak_field}.

Two factors increase the chances of photon scattering by vacuum becoming more noticeable than generation from gas: a decrease in pressure and an increase in the intensity of the incident pulse. This point is especially significant because our calculations show that, in helium, the growth of scattered-photon intensity with incident-pulse intensity is markedly slower than in vacuum.
This does not contradict the fact that for a given intensity $I_0$ of interacting laser pulses, the maximum intensity of the scattered field increases in proportion to the square of the pressure. For example, at peak laser intensity $I_0=10^{22}$~W/cm$^2$ we need a fairly low pressure, such as $P\approx 10^{-12}$~Torr, so that the contribution of vacuum and gas  WMP processes to be comparable for the considered harmonics, see Fig.~\ref{mono_spec}~(b). By increasing the peak intensity of incident pulses to $I_0=10^{23}$ W/cm$^2$, we achieve equality of the peak intensities of generated harmonics in He and vacuum at a pressure 30 times higher, see Fig.~\ref{mono_spec}(d). Pressure $P = 3\cdot10^{-11}$~Torr, corresponding to this most optimistic case, can be achieved on a modern experimental setup.

\begin{figure}[t]
	\centering{
		\includegraphics[scale=0.42,trim={0 0 0 0},clip]{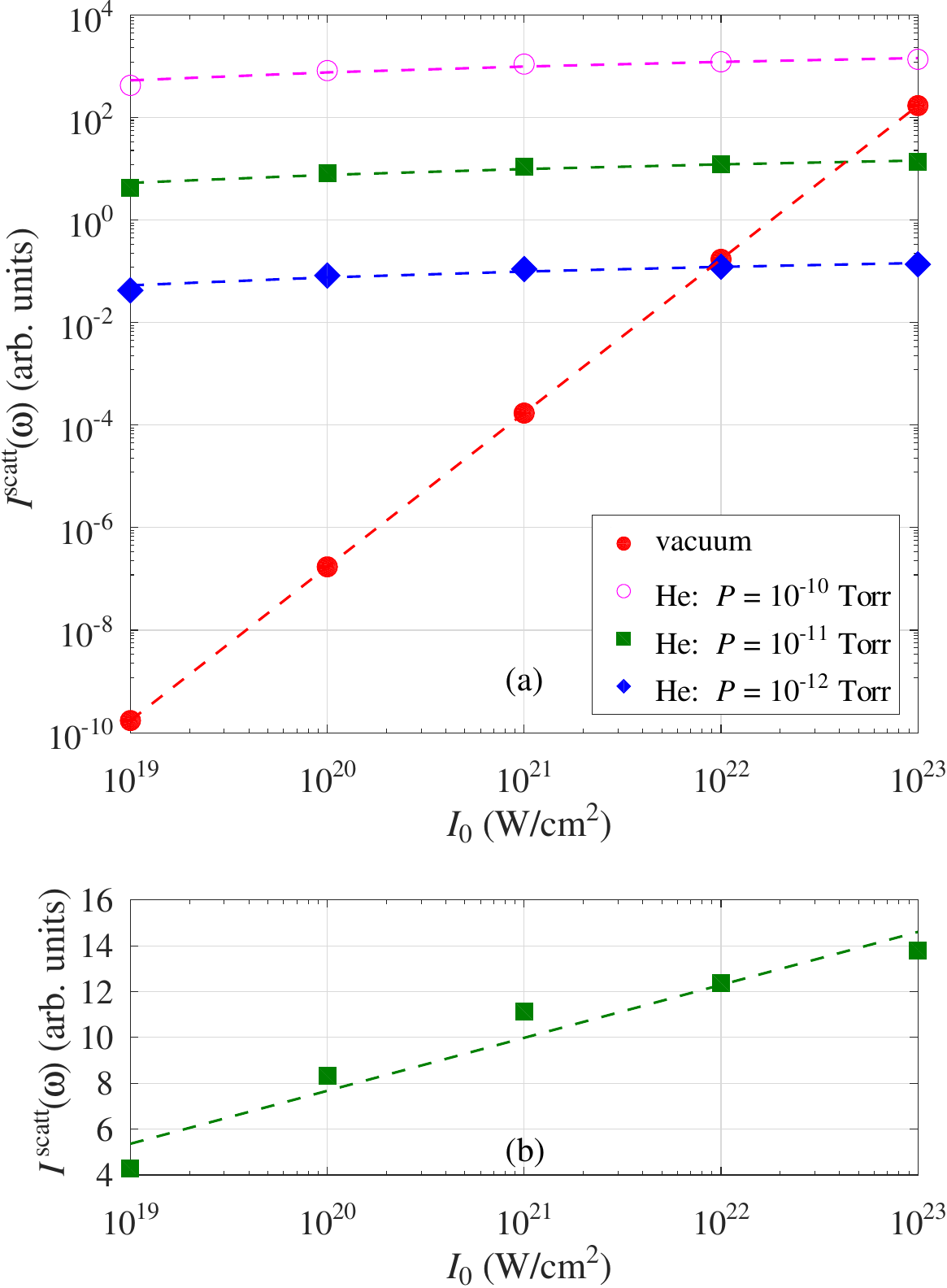}}
	\caption{The maximum spectral intensity of the scattered field as a function of the peak intensity of the incident pulses in the form \eqref{E_one_mono}. (a) Dependencies are shown for vacuum and for He with different initial pressures  in log-log axes. Dashed lines are used to model linear (in vacuum) and logarithmic (in He) regression of the numerical results. (b) The same dependence for He at a pressure of $P=10^{-11}$ Torr on a semi-logarithmic scale. The dotted line represents a linear regression simulation $I^{\mathrm{scatt}}=2.31\cdot~\log_{10}I_0~-~38.52$ of the numerical results.} \label{vacVSplas}
\end{figure}

Figure~\ref{vacVSplas}(a) shows the simulated dependencies between the intensities of the incident laser pulses and the maximal intensity of scattered photons in vacuum and gas. For vacuum, the linear regression analysis of our numerical results yields a slope of 3.0001 on a log-log scale, which is consistent with the idea of a dominant third-order nonlinear process \cite{Moulin99}. However, for WMP in gas, although the intensity of the scattered field increases with the intensity of the incident pulses, it occurs much more slowly: instead of a cubic law, it is closer to a logarithmic one. Moreover, if we have a look on this dependence for gas in log-linear scale, see Figure~\ref{vacVSplas}(b), we can observe that the growth is not even clearly linear, although we can find the corresponding regression coefficients. It is worth to note that the logarithmic dependence $I^\mathrm{scatt}(I_0)$ was also obtained earlier in unpublished simulation results by G. B\'elanger, C.Lefebvre, F.Fillion-Gourdeau and S.~MacLean (2017) when studying WMP in a much denser gas ($10^{24}$~m$^{-3}$) for $I_0\in[10^{14}, 10^{24}]$~W/cm$^{2}$ within a PIC-based model. 
Indeed, in the ultra-high intensity regime of interest to us, the expressions for the polarization and magnetization of the vacuum are proportional to the third power of the electromagnetic field (with higher order corrections), see \eqref{PxPz}, \eqref{My}, however, in the same regime, the third-order susceptibility approximation becomes inapplicable to fully ionized relativistic plasma, and we need to solve the system~\eqref{tranE2}-\eqref{current2_z}, which shows a slower increase with $I_0 $.
In practice, differences in scaling laws could be used to determine whether WMP is generated by vacuum or by particle processes by performing spectrum measurements at different laser intensities. Numerical confirmation, whether similar scaling holds in 3D is still an open problem. We add our qualitative reasoning on this issue in the Conclusion section.

\subsection{Generation of WMP photons in the interaction of two broadband laser pulses}\label{broad}
\begin{figure*}
	\centering{
		\includegraphics[scale=0.9,trim={0 0 0 0},clip]{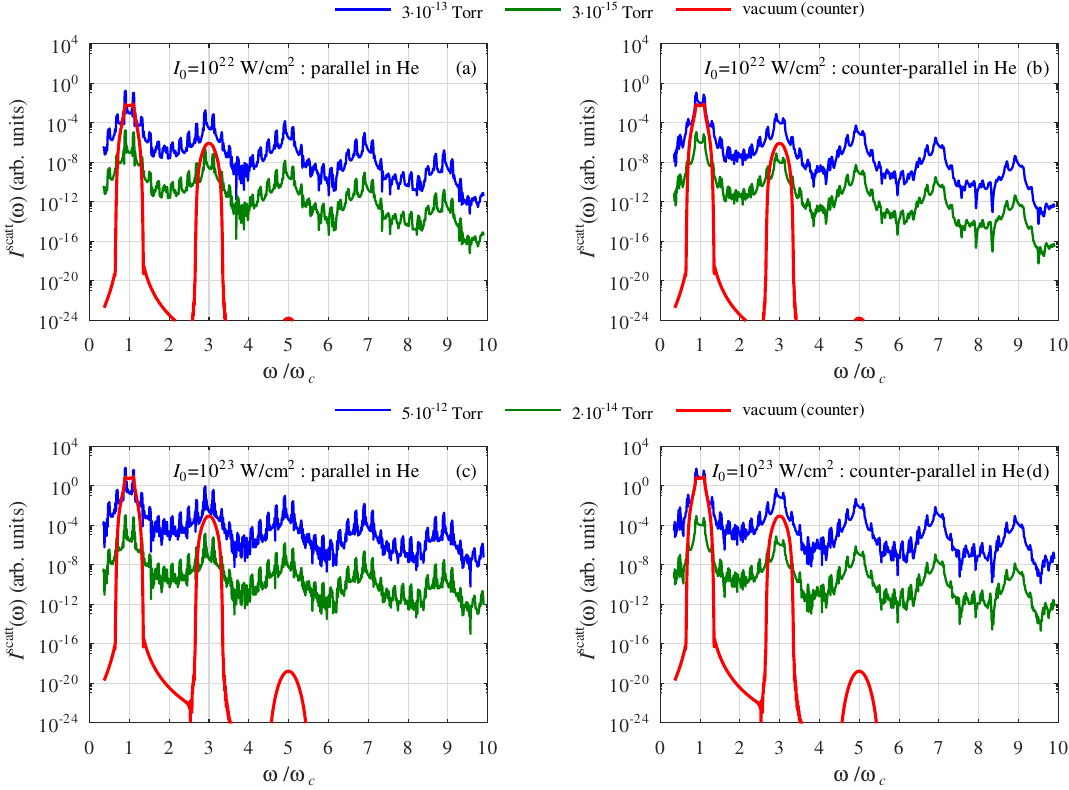}}
	\caption{Spectra of the scattered field generated by the interaction of two envelopes given by formula~\eqref{E_one_band} with wavelengths in range $\lambda\in[720, 880]$ nm and intensities $I_0$ either (a),(b) $10^{22}$ W/cm$^2$ or (c),(d) $10^{23}$ W/cm$^2$.
	Solutions in helium are shown in inserts (a),(b) for pressures $P=3\cdot10^{-13}$ Torr - in blue and $P=3\cdot10^{-15}$ Torr - in green, and in inserts (c),(d) for $P=5\cdot10^{-12}$ Torr - in blue and $P=2\cdot10^{-14}$ Torr - in green; inserts (a),(c) correspond to co-propagatin and (b),(d) to counter-propagating pulses in helium. For comparison, vacuum solutions are shown in red for the same envelopes, the same intensities, but always for counter-propagating pulses. The frequency axis is normalized to central frequency $\omega_c=\pi c(\lambda_\mathrm{min}+\lambda_\mathrm{max})/\lambda_\mathrm{min}\lambda_\mathrm{max}$.} \label{broadband_spec}
\end{figure*}

Now we compare the harmonics resulting from the interaction of two envelopes of the form \eqref{E_one_band}, also called case (B) in Section~\ref{num_exp}, see Fig.~\ref{broadband_spec} (a)-(d). 
This time we do not see much difference between the co- and counter-propagation in helium, as was noticed in the case of interaction of envelopes of type (A), compare with Fig.~\ref{mono_spec}. Another feature is that with the same central frequency $\omega_c$ of both envelopes in centrosymmetric media, maxima in the spectra appear only at odd harmonics $(2n+1)\omega_c$, where in our numerical simulations  $n=0, 1, \dots 5$. 
In addition, the peaks are noticeably wider and have a substructure - spikes in the spectra, separated by a distance $(\omega_\mathrm{max}-\omega_\mathrm{min})/ \omega_c=0.2$ in relative units. 

Again, we see that if in vacuum the increase in the intensity of generated photons is three orders of magnitude per one order of magnitude of the incident pulse, then for gas this increase is less than one order of magnitude, retaining the pattern qualitatively similar to that shown in Figure~\ref{vacVSplas}. Although in this case the quantitative equality between the spectral harmonics of different origin is achieved at lower pressure. 

\begin{figure}[t]
	\centering{
		\includegraphics[scale=0.42,trim={0 0 0 0},clip]{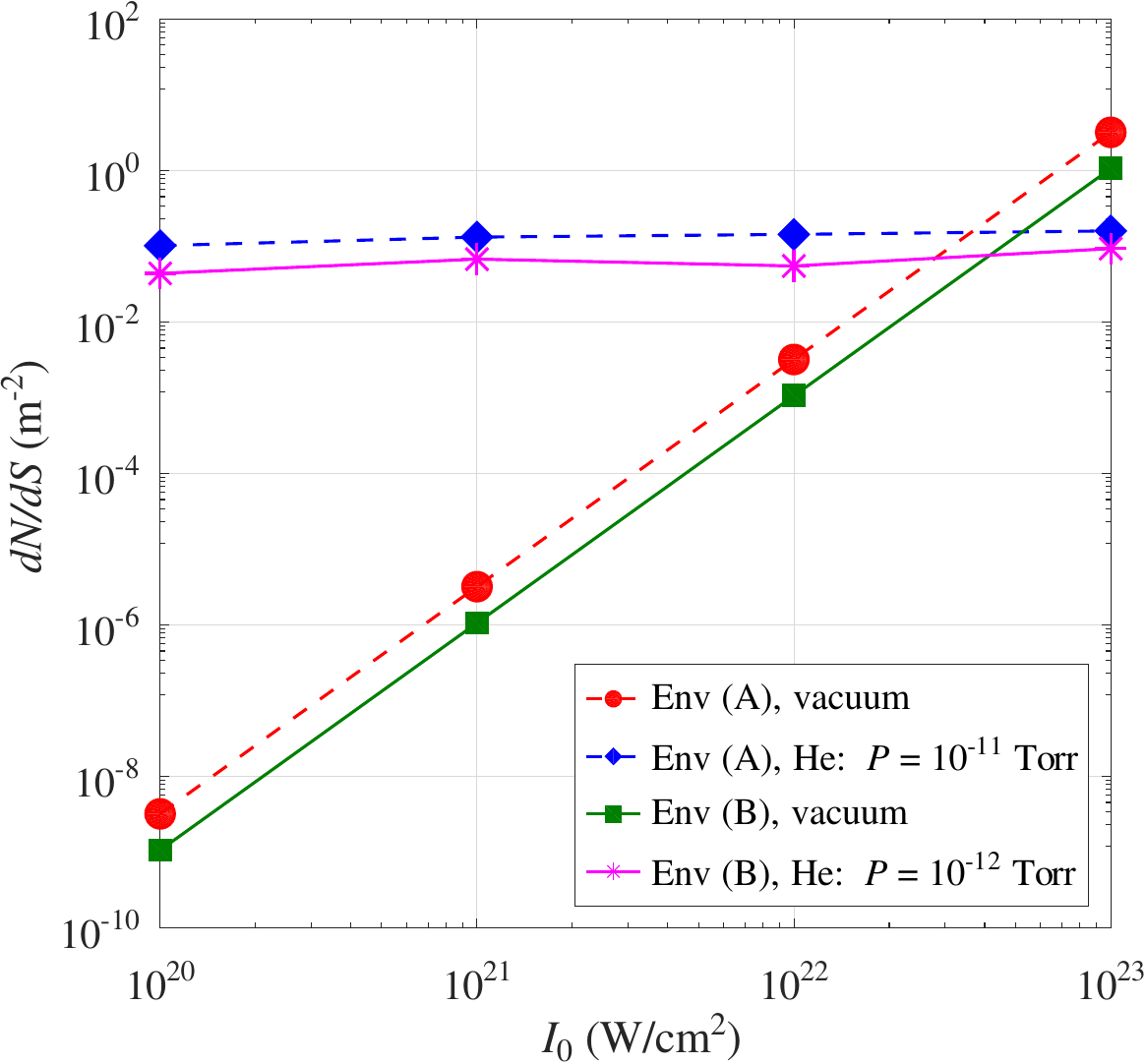}}
	\caption{The total number of photons generated per element of a solid surface as a function of the peak intensity $I_0$ of incident pulses with envelopes of the form ``Env (A)'' defined by formulas \eqref{E_one_mono} and ``Env (B)'' - of the form \eqref{E_one_band}.} \label{num_phot}
\end{figure}

Even in modern experiments it is difficult to measure the WMP photon spectra, there are too few of them, so a more convenient parameter is the total number of generated photons, which we calculate from \eqref{Poynting} for gas, or using a similar formula for vacuum, and then convert to common units of measure. The result is presented in Figure~\ref{num_phot}, where we compare numbers of generated photons per solid surface element for incident pulses \eqref{E_one_mono} and \eqref{E_one_band}. According to our simulations, in vacuum, the interaction of pulses with different and distant central frequencies produces more photons than the interaction of two broadband trains with the same central frequencies. On the other hand, in gas at a fixed pressure, the total number of photons is higher when broadband trains interact, which means that a lower pressure is required to balance this number with the number of photons generated in vacuum. As follows from Figure~\ref{num_phot}, in case of broadband pulse interaction, at pressure $10^{-12}$ Torr the number of WMP-photons from vacuum begin to exceed the number of WMP-photons from plasma at $I_0=4\cdot10^{22}$ W/cm$^2$. By applying the patterns studied, to expect the parity of contributions from vacuum and gas, an increase in pressure by $\approx\sqrt{10}$ times up to $\approx3\cdot10^{-12}$ Torr requires an increase in pulse intensities up to $10^{23}$ W/cm$^2$. As we mentioned in the introduction, this pressure can be achieved with modern devices. Moreover, directly in the region of interaction, the pressure can be even lower due to the fact that the incoming pulse sweeps particles from the optical axis with its relatively low-energy front.   

\section{Conclusion}\label{cncls}
We simulated WMP processes in gas and vacuum to find the conditions under which the generation of photons in both media has the same level. In the case of gas, to reduce the numerical noise when solving the 1D-1V Vlasov equation, we applied a semi-Lagrangian method with parallelization to speed up the simulations. For the vacuum, we used the semi-classical weak-field expansion for vacuum polarization and magnetization.

We have found that the intensity of generated photons in vacuum scales with increasing peak intensity of the incident pulses as a third order process, whereas when photons are generated in a relativistic ionized gas, their intensities increase only at a logarithmic rate.
We have concluded that for equal output efficiency of generated photons at peak intensities of about $10^{23}$ W/cm$^2$, a pressure of $10^{-11} - 10^{-12}$~Torr is required - depending on the characteristics of the interacting pulses. Such pressures can be achieved with relatively standard ultra-high vacuum techniques. Moreover, as recent experiments and simulations have demonstrated, the electron expulsion due to the ponderomotive force of the pedestal and the main pulse (ponderomotive explosion) \cite{ravichandran2025} can play an important role in creating a near-perfect vacuum. By thus reducing the effective electron pressure, this effect should work in favor of QED processes compared to wave mixing in plasma.

It should be also noted that our analysis only considers the interaction region, where the field and nonlinear effect are maximal. In a real experimental setup where high intensities are reached via tightly focused fields, one should also take into account the propagation of the laser before in reaches the focus. This propagation occurs on longer length scales, albeit much lower intensity, so could have an important contribution in number of generated photons. Evaluating this contribution numerically is a complex multiscale problem, outside the scope of this article. However, together with the effect of pressure reduction by the front of the laser pulse, this makes our estimate of the required pressure the minimum limit,  i.e. the chances of detecting WMP photons may be even higher at the proposed pressure.

Radiation-reaction was not taken into account in the rarefied gas, but its effect will start to be significant at  $I_0=10^{23}$ W/cm$^2$ and above. This would also lead to losses of electron energy and, consequently, to an amplitude reduction of generated harmonics in the helium plasma in favor of the QED process.

The generalization to higher dimensions (2D and 3D) is not straightforward. Indeed, it is well-known that dimensionality has an important influence on certain particle acceleration processes in PIC simulations. It is generally accepted that higher dimension simulations produce particles with lower energies (see Refs. \cite{doi:10.1063/1.4982741,PhysRevE.104.045206} for examples). This correction, which compensates for the overestimation of currents generating scattered photons in gas or plasma, also increases the chances of detecting WMP as the scattering of light-by-light. To what extent such dimensionality effects can actually increase our pressure estimate will be the subject of further research.

\appendix

\section{Computational algorithm for gas}\label{plasma_alg}

Here we describe the computational algorithm for the system of equations \eqref{Ampere2}-\eqref{current2}. In our scheme, the following variables propagate in time: transverse $E^\pm_{\mathrm{scatt}}(z,t)$ and longitudinal $E_z(z,t)$ electromagnetic fields, and the distribution functions $f_s(z,p_z,t)$ ($s = \{$e$^{-}$, He$^{2+}\}$), whereas the momentum $p_x(z,t)$ and currents $J_x(z ,t )$, $J_z(z,t)$ are calculated through these variables at the required time nodes.

This is how the elementary block of the numerical scheme, repeating in time, looks like. If $n>0$, we start with advecting the transverse scattered fields from $t_{n-1/2}$ to $t_n^-$ (time ``just before'' $t_n$)
\begin{equation}\label{Epm_free1_num}
E^\pm_{\mathrm{scatt}}(z_i,t_{n}^-)=E^\pm_{\mathrm{scatt}}\Big(z_i\mp \dfrac{\Delta t}{2},t_{n-1/2}\Big),
\end{equation}
or, if $n=0$, we plug in the initial conditions $E^\pm_{\mathrm{scatt}}(z_i,t=~0) = 0$. Using the analytic expression \eqref{E_one_mono} or \eqref{E_one_band} for the laser field at $t_n$, we calculate the magnetic field $B_y(z_i, t_n)$ according to \eqref{b_tot}. Then we can numerically solve \eqref{px_evol_norm} in space using the midpoint method or any multi-step method such as the Adam-Moulton method \cite{Quart}.

Now we can compute the currents of charged particles at the point $t_n$  using the value of $p_x (z_i, t_n)$ and the distribution function $f_s(z_i, (p_z)_j, t_n)$ calculated in the previous block of the scheme (or known from the initial conditions if $n=0$):
\begin{align}
\label{curr_x_num}
\nonumber J_x\!(z_i,t_{n})\!&=\! \sum_s\!\dfrac{Q_sp_x(z_i,\!t_{n})}{M_s}\!\sum_j\! \dfrac{f_s(z_i, (p_z)_j, t_{n})}{\gamma(p_x(z_i,t_{n}),\!(p_z)_j,\!M_s)}\Delta p_z,\\
\\
J_z\!(z_i,t_{n})\!&=\! \sum_s\dfrac{Q_s}{M_s}\sum_j \dfrac{f_s(z_i, (p_z)_j, t_{n})(p_z)_j}{\gamma(p_x(z_i,t_{n}),(p_z)_j,M_s)}\Delta p_z.
\end{align}
We use these currents to ``kick'' the transverse and longitudinal fields in time:
\begin{align}
E^\pm_{\mathrm{scatt}}(z_i,t_{n}^+)&=E^\pm_{\mathrm{scatt}}(z_i,t_{n}^-)-J_x(z_i, t_{n})\Delta t,\\
E_z(z_i,t_{n+1/2})&=E_z(z_i,t_{n-1/2})-J_z(z_i, t_{n})\Delta t.
\end{align}
The last step of the algorithm for the transverse field is again the advection half step:
\begin{equation}\label{Epm_free2_num}
E^\pm_{\mathrm{scatt}}(z_i,t_{n+1/2})=E^\pm_{\mathrm{scatt}}\Big(z_i\mp \dfrac{\Delta t}{2},t^+_{n}\Big).
\end{equation}
To solve the kinetic equation in a highly relativistic regime, we cannot use the time splitting scheme common for a laser beam of lower energies \cite{Ghizzo90}, since this leads to an accumulation of error in time \cite{Huot}. Instead, as in the last mentioned article, we implement full 1d-1d advection for a simultaneous shift in space and time:
\begin{eqnarray}\label{f_2dadvect}
f_s(z_i,(p_z)_j, t_{n+1}) = f_s(z_d,(p_z)_d, t_{n}),
\end{eqnarray}
where the departure points $(z_d,(p_z)_d)$ of the trajectories are to be found iteratively using the midpoint rule \cite{Durran,fletcher2019semi}, so that for every iteration step $k>0$:
\begin{align}\label{z_d}
\nonumber &z^{(k)}_d = z_i-\Delta t\dfrac{(p_z)^{(k-1)}_{mid}}{M_s\gamma_s(p_x(z^{(k-1)}_{mid},t_{n+1/2}),(p_z)^{(k-1)}_{mid},M_s)}, \\
\\
\nonumber & (p_z)_d^{(k)}=(p_z)_j-\Delta tQ_s\Big[E_z(z^{(k-1)}_{mid},t_{n+1/2})+\\
\label{zp_d}
\nonumber&\dfrac{p_x(z^{(k-1)}_{mid},t_{n+1/2})}{M_s\!\gamma_s(p_x(z^{(k-1)}_{mid}\!,\!t_{n+1/2}),(p_z)^{(k-1)}_{mid}\!,\!M_s)}\!B_y(z^{(k-1)}_{mid}\!,\!t_{n+1/2})\Big],\\
\end{align}
where denoted
\begin{eqnarray}
z^{(k)}_{mid} &=& \dfrac{z_i+z_d^{(k)}}{2}, \\
(p_z)^{(k)}_{mid}&=& \dfrac{(p_z)_j+(p_z)_d^{(k)}}{2},
\end{eqnarray}
and iterations start with
\begin{align}
&z^{(0)}_d = z_i-\Delta t\dfrac{(p_z)_j}{M_s\gamma_s(p_x(z_i,t_{n+1/2}),(p_z)_j,M_s)},\\
\nonumber& (p_z)_d^{(0)}=(p_z)_j-\Delta tQ_s\Big[E_z(z_i,t_{n+1/2})+\\
&\;\dfrac{p_x(z_i,t_{n+1/2})}{M_s\gamma_s(p_x(z_i,t_{n+1/2}),(p_z)_j,M_s)}B_y(z_i,t_{n+1/2})\Big].\qquad
\end{align}
Note that to complete this last step of the block, we first need to calculate $p_x$ now at time $t_{n+1/2}$.

In the semi-Lagrangian approach for the Vlasov-Maxwell system, some interpolation techniques are usually used to implement the equations \eqref{Epm_free1_num}, \eqref{Epm_free2_num}-\eqref{zp_d}. Namely, we use cubic spline interpolation in the $z$-direction to solve the field advection equations \eqref{Epm_free1_num} and \eqref{Epm_free2_num}, as well as for finding the ``departure'' points $(z_d, (p_z)_d)$ in iteration cycle \eqref{z_d}, \eqref{zp_d}, where we need to compute $p_x$, $E_z$ and $B_y$ in between the mesh nodes.

A more computationally expensive problem is the two-dimensional interpolation required to compute advection in \eqref{f_2dadvect}. For this task, we used bilinear interpolation, which is fast but can potentially reduce the accuracy of the results. To check this, we numerically tested the order of convergence of the entire computational model. Namely, for the distribution functions, we obtained the order of convergence 2.5 and 2 for the variables $z$ and $p_z$, respectively, but only 1.6 for $t$. In addition to the numerical measurement of the rate of convergence, we also use the norm of the distribution function to estimate the accuracy and stability of the scheme. Thus, in our calculations, the variations in the norm at the relevant time steps were within $2\%$. Moreover, for fields, the convergence order turned out to be 2 for both variables, $z$ and $t$. Since from a practical point of view we are most interested in the fields/spectra, we can conclude that our model provides a fairly good convergence (theoretically it should be 2). In addition to the speed of calculating bilinear interpolation, it is also important for us because our computational model uses parallelization, which means that the interpolation must be local for each subdomain. 

\begin{figure}[b]
	\centering{
		\includegraphics[scale=0.42,trim={0 0 0 0},clip]{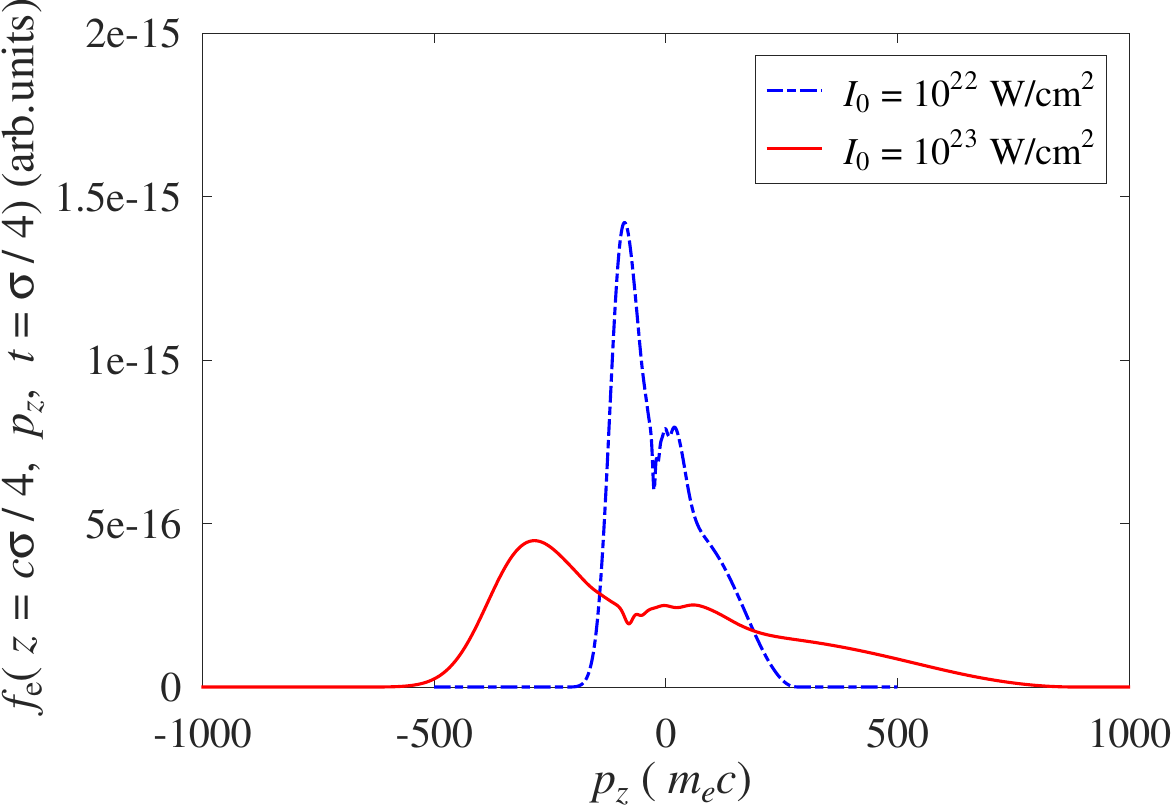}}
	\caption{Distribution function of the electrons in momentum $p_z$ at fixed space position $z$ and time $t$ for different incident intensities of the interacting pulses.} \label{distr}
\end{figure}

Before describing domain decomposition, we discuss step sizes for all variables. In our simulations $\Delta z\approx0.05$ (in units $\lambda_1/2\pi$) and $\Delta p_z\approx0.2-0.25$ (in units $m_\mathrm{e}c$), which provide accurate description of evolving electromagnetic waves and distribution functions. The spatial size of the phase space in our simulations is $10c\sigma$ and does not change, but we need to increase the limits of the momentum region when we increase the values of peak intensities in the calculations. Thus, if for $I_0=10^{22}$ W/cm$^2$ we consider $p_z\in [-500, 500] m_ec$, but for $I_0=10^{23}$ W/cm$^2$ we should increase these limits up to $p_z\in [-1000, 1000] m_ec$, see Figure~\ref{distr}. To estimate the acceptable time step $\Delta t$ one should use the deformational Courant number like in \cite{Grizzo03}, and as follows from \eqref{z_d} and \eqref{zp_d}, the time step should decrease inversely proportional to the increase in the laser field amplitude. For example, we use $\Delta t = 0.006$ (in units $\lambda_1/2\pi c$) for peak intensity $I_0=10^{22}$ W/cm$^2$ and $\Delta t = 0.0015$ for $I_0=10^{23}$ W/cm$^2$. 

\begin{figure}[t]
	\centering{
		\includegraphics[scale=0.9,trim={0 0 0 5},clip]{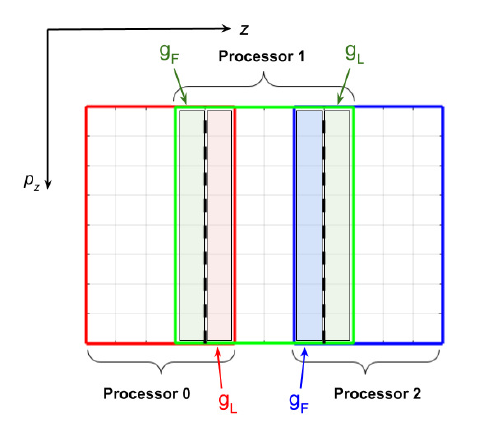}}
	\caption{Domain decomposition for parallelization on the example of 3 processors, see the text. Domain boundaries are shown as thick lines of different colors, dashed lines show the boundary between subdomains if there were no ghost rows.}\label{DD}
\end{figure}

In our simulations, for phase space ($z, p_z$) we use grids size up to 32768$\times$8192 (or 209$ \mu$m$\times2048$ $m_\mathrm{e}c$), and hundreds of thousands steps in time, resulting in long computation times. Fortunately, the problem can be parallelized. For the steps sizes defined above, using the relation \eqref{z_d}, we can conclude that in one time step the $z$-coordinate along the trajectory in the phase space should be located in no more than two adjacent cells of the space. For comparison, in the case of the $p_z$ coordinate, see \eqref{zp_d}, the jump during $\Delta t$ can span more than 6 cells. We use this fact to perform the domain decomposition of our phase space. Namely, we slice our domain across the $z$-axis dividing it in $N_p$ regions by the number of processors. Further, each such a subdomain, with the exception of the initial and final ones, we supplement with two 	``ghost''-rows: first ($g_F$) and last ($g_L$). The initial domain is supplemented with the last ghost and the final - with the first ghost only, see an example in Fig.~\ref{DD}. 
Now we can solve the Vlasov equations for each subdomain separately, the only thing we need to take care of is to update the information in the ghost rows with information from the neighboring processors after each time step. 
Note that in this model, we interpolate the distribution function for each subdomain separately, which in itself gives a large gain in time compared to interpolation over the entire area. The necessary condition here is the use of local interpolation. Another part of algorithm, which can be accelerated with MPI is computing currents $J_x$, $J_z$ in subdomains and then gathering them to calculate the fields. We compute $E_x(z,t), B_y(z,t), E_z(z,t)$ and momentum $p_x(z,t)$ on the root processor and use 1d cubic spline interpolation (nonlocal) for them. It makes no gain to parallelize the calculation of the field variables, since they are already executed more than 5,000 times faster than modeling of the distribution function. It is noteworthy that the speedup factor (relative to the use of only one processor) due to such parallelization can be even greater than the number of processors used, probably due to the fact that the 2d interpolation efficiency grows faster than the area of the interpolation region decreases.

\section{Computational algorithm for vacuum}\label{vacuum_alg}
The numerical scheme that we use to study WMP in vacuum is similar to that for gas, however this time we can use the analytical expressions for currents \eqref{Jx} and \eqref{Jz}. It can be seen from the expression \eqref{Jz} for $J_z$ that this component is non-zero only in the case of a non-zero field $E_z$. However, if we turn to the equation $\nabla~\cdot~{\bf E}=-\nabla\cdot{\bf P}$ with ${\bf P}$ from \eqref{PxPz}, we see that in the chosen geometry the longitudinal field is always zero. In principle, the situation can be changed by adding a static electric field, but here our goal is to compare the results with the WMP in gas, where we do not assume that there is any external electric field. That is why in our simulations we can skip the equation for $E_z$ at all.
 
As in case of gas, we rewrite the first two equations in \eqref{Max_mod} for transverse field $E^\pm = E_x\pm B_y$ as
\begin{equation}
(\partial_t\pm\partial_z)E^\pm = - J_x.
\end{equation}
Further we describe one time block of the numerical scheme. If $n>0$, we start with advecting the scattered transverse vacuum fields from $t_n$ to $t_{n+1/2}^-$ (time ``just before'' $t_{n+1/2}$):
\begin{equation}\label{Epm_vac1_num}
E^\pm_{\mathrm{scatt}}(z_i,t_{n+1/2}^-)=E^\pm_{\mathrm{scatt}}\Big(z_i\mp \dfrac{\Delta t}{2},t_n\Big),
\end{equation} 
or, if $n=0$, we plug in the initial conditions $E^\pm_{\mathrm{scatt}}(z_i,t=~0) = 0$.

Now, using the analytic expression for the laser field at $t_{n+1/2}$, we calculate \eqref{e_tot}, \eqref{b_tot} and then the current $J_x(z_i, t_{n+1/2})$ for which we use central differences when estimating the spatial derivatives $\partial_zE_x$, $\partial_zE_z$, $\partial_zB_y$. We use this current to ``kick'' the transverse scattered fields in time:
\begin{eqnarray}
\nonumber E^\pm_{\mathrm{scatt}}(z_i,t_{n+1/2}^+)&=&E^\pm_{\mathrm{scatt}}(z_i,t_{n+1/2}^-)-J_x(z_i, t_{n+1/2})\Delta t,\\
\end{eqnarray}
The last step of the algorithm for the transverse field is again the advection half step:
\begin{equation}\label{Epm_vac2_num}
E^\pm_{\mathrm{scatt}}(z_i,t_{n+1})=E^\pm_{\mathrm{scatt}}\Big(z_i\mp \dfrac{\Delta t}{2},t^+_{n+1/2}\Big).
\end{equation}
As in the case of gas, for steps \eqref{Epm_vac1_num} and \eqref{Epm_vac2_num} we use cubic spline interpolation.

The numerical order of convergence for this scheme is found to be 2 in $z$ and $t$, and the simulation uses the same spatial and temporal steps as in the gas case.

\begin{acknowledgments}
The authors would like to thank Catherine Lefebvre for sharing previously unpublished results of WMP simulations in gases obtained using the PIC code and St\'ephane Payeur for useful discussion.
This research was enabled in part by support provided by the Digital Research Alliance of Canada \url{alliance​can​.ca}. 
\end{acknowledgments}

\bibliography{biblio}

\end{document}